\documentclass[epj]{svjour}
\usepackage{graphics}
\usepackage{adjustbox}
\usepackage{sidecap}
\begin{document}
\bibliographystyle{ieeetr}

\title{Scaling Properties in Time-Varying Networks with Memory}
\subtitle{}
\author{Hyewon Kim\inst{1} \and Meesoon Ha\inst{2}\thanks{\emph{Correspondence to:} msha@chosun.ac.kr}%
\and Hawoong Jeong\inst{3}\thanks{hjeong@kaist.edu}
}                     
%
%
\institute{Department of Physics, Korea Advanced Institute of Science and Technology, Daejeon 34141, Korea \and Department of Physics Education, Chosun University, Gwangju 61452, Korea \and Department of Physics and Institute for the BioCentury, Korea Advanced Institute of Science and Technology, Daejeon 34141, Korea}
\date{Received: date / Revised version: date}
%
\abstract{
The formation of network structure is mainly influenced by an individual node's activity and its memory, where activity can usually be interpreted as the individual inherent property and memory can be represented by the interaction strength between nodes. In our study, we define the activity through the appearance pattern in the time-aggregated network representation, and quantify the memory through the contact pattern of empirical temporal networks. To address the role of activity and memory in epidemics on time-varying networks, we propose temporal-pattern coarsening of activity-driven growing networks with memory. In particular, we focus on the relation between time-scale coarsening and spreading dynamics in the context of dynamic scaling and finite-size scaling. Finally, we discuss the universality issue of spreading dynamics on time-varying networks for various memory-causality tests.
} 
\maketitle
\section{Introduction}
\label{intro}

Network analysis becomes essential nowadays to figure out the dynamical and structural properties of interacting systems, such as physical, biological, social, and economic systems. Most network studies have been done in static representations, where network structures are almost fixed for the simplicity. However, in general, interactions among individuals in the system are {\it time-varying} and links are also of short duration~\cite{Holme2012}. As a result, the system forms a so-called ``{temporal network}"~\cite{Holme2015}, where structural changes of the system in time cannot be ignored anymore. For the case of temporal networks~\cite{Scholtes2014,Holme2005,Gautreau2009}, unlike the case of static ones, it is very important to consider both structural properties and time-ordered records of events. Therefore, in dynamical systems, understanding the network structure formation in time becomes crucial.

Along with the development of technologies, it has been possible to obtain time-stamped data~\cite{Eagle2006,Cattuto2010,Kossinets2006,Stehle2011,Barrat2013,Barrat2014}. 
Lots of efforts to understand dynamical systems have been tried with various methods, from empirical data analyses to analytical approaches~\cite{Scholtes2014,Rocha2011,Karsai2011,Starnini2012,Lentz2013,Masuda2013,Valdano2015}, which indicate that the time-ordered information of links and resulting network topologies do alter the properties of spreading phenomena in epidemics. 
Previous studies were mostly based on connectivity-driven models that properly describe structural features with very slow changes, but are not appropriate for short-time scale properties. To overcome such limitations, various time-scale tests 
have been discussed in~\cite{Barabasi2005,Rosvall2010,Caceres2011,Min2013,Perotti2014} with ambiguous definitions. Moreover, time-window graphs have also been studied to answer the question of finding the optimal time-window size of the system, of which more detailed discussions can be found in~\cite{Caceres2011} and the most recent review paper~\cite{Holme2015}~(see references therein for the Sect. 3.2.3 also).
On the other hand, a new model called activity-driven model were suggested~\cite{Perra2012} that can generate the instantaneous time description of the networks. While Markovian activity-driven models for human activity patterns can easily explain the process of structural features in time-varying networks, they cannot capture the existence of memory (either non-Markovian or history-dependent time-causality), which plays an crucial role in emerging dynamical
heterogeneities~\cite{Karsai2011,Dorogovtsev2000,Karsai2012,Karsai2014,Vestergaard2014,Moinet2015}. For a better discussion on the origin of such anomalous behaviors due to the memory effect, several attempts have been performed, where memory is considered as exploration~\cite{Karsai2014}, preferential return~\cite{Song2010}, social reinforcement~\cite{Laurent2015}, and triadic-close mechanism~\cite{Memory-JSTAT2014}, respectively.

\begin{figure}[]
\center
\includegraphics*[width=0.95\columnwidth]{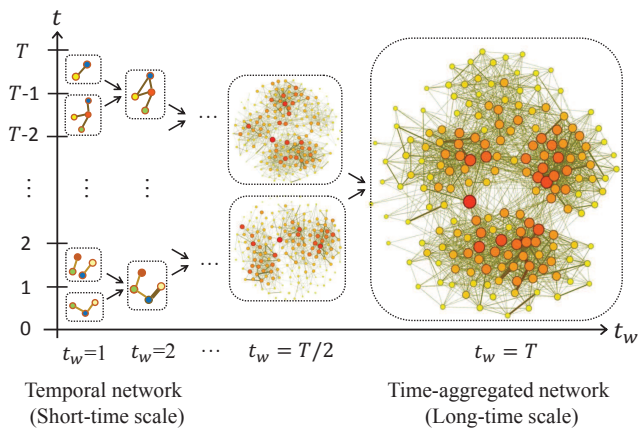}
\caption{(Color online) Temporal networks are schematically illustrated, where $t_w$ is the window size of observable time, and $T$ is the maximum time. Here we use the empirical dataset of School (2012)~\cite{SocioWeb} as network realizations. Note that the difference in colors and symbol sizes represents the difference in nodes' degrees and strengths.}
\label{fig1-snapshot}       
\end{figure}

In this paper, we propose a modified activity-driven model with the memory effect on temporal networks, which can be understood as a simple variant of previous models. Using the time-aggregated networks of five real datasets~\cite{SocioWeb,Brazil-Rocha2010,Email-Ebel2002}, we estimate the range of network parameters to provide the proper definitions of activity and memory in our model. After tuning the parameters, we investigate the role of activity and memory in epidemics~\cite{EP-RMP2015} (by the conventional susceptible-infected model, SI), temporal-pattern coarsening in time-varying networks (by a random greedy walker, similar to the most recent study~\cite{SHolme2015}), and aging tests in coarsening schemes, where we employ the extended finite-size scaling (FSS) technique against the time-window size ($1\le t_w\le T$) as illustrated in Figure~\ref{fig1-snapshot}. Finally, we conclude the paper with the summary of main results and some remarks as well as open questions for future studies. 

\section{Model}
\label{model}

As key features in human social behaviors, {e.g.}, epidemic spreading~\cite{Vespignani2009,Eubank2004,Ginsberg2009}, we consider both activity and memory of individuals. While the former can be interpreted as the personal inherent ability, the latter can be understood as the personal information for the internal record of previous contacts as time elapses. Hence memory can be treated as a physical quantity that is acquired through social environment, which is first encoded, kept stored with infinite capacity, and retrieved in future dynamics. Particularly, in our model, memory is used as the {\it time-accumulated} personal (long-term) experience, {i.e.}, the non-Markovian strength with the individual history and preference per node. Based on the similar concept, previous studies have used a variety of definitions: activity potential~\cite{Perra2012}, reinforced process~\cite{Karsai2014}, the interplay of exploration and preferential return~\cite{Song2010}, and triadic-closure mechanism~\cite{Memory-JSTAT2014}. 

Based on the network analysis of five different empirical datasets: School~(2011,~'12), Hospital~\cite{SocioWeb}, Brazil~\cite{Brazil-Rocha2010}, and Email~\cite{Email-Ebel2002}, 
in the context of time-aggregated networks, we propose a modified activity-driven model with the memory effect. Particularly, we tune model parameters to understand the temporal patterns of empirical datasets. 
\begin{figure}[]
\center
\includegraphics[width=0.75\columnwidth]{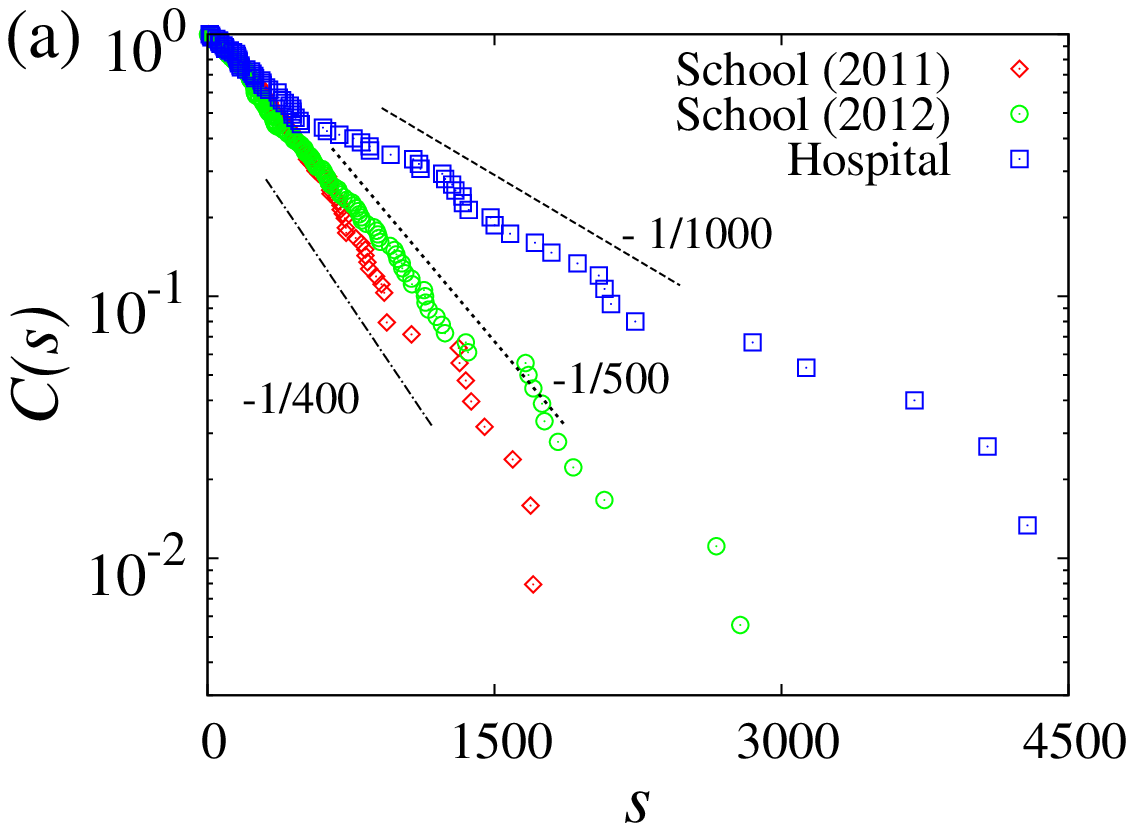}\\
\includegraphics[width=0.75\columnwidth]{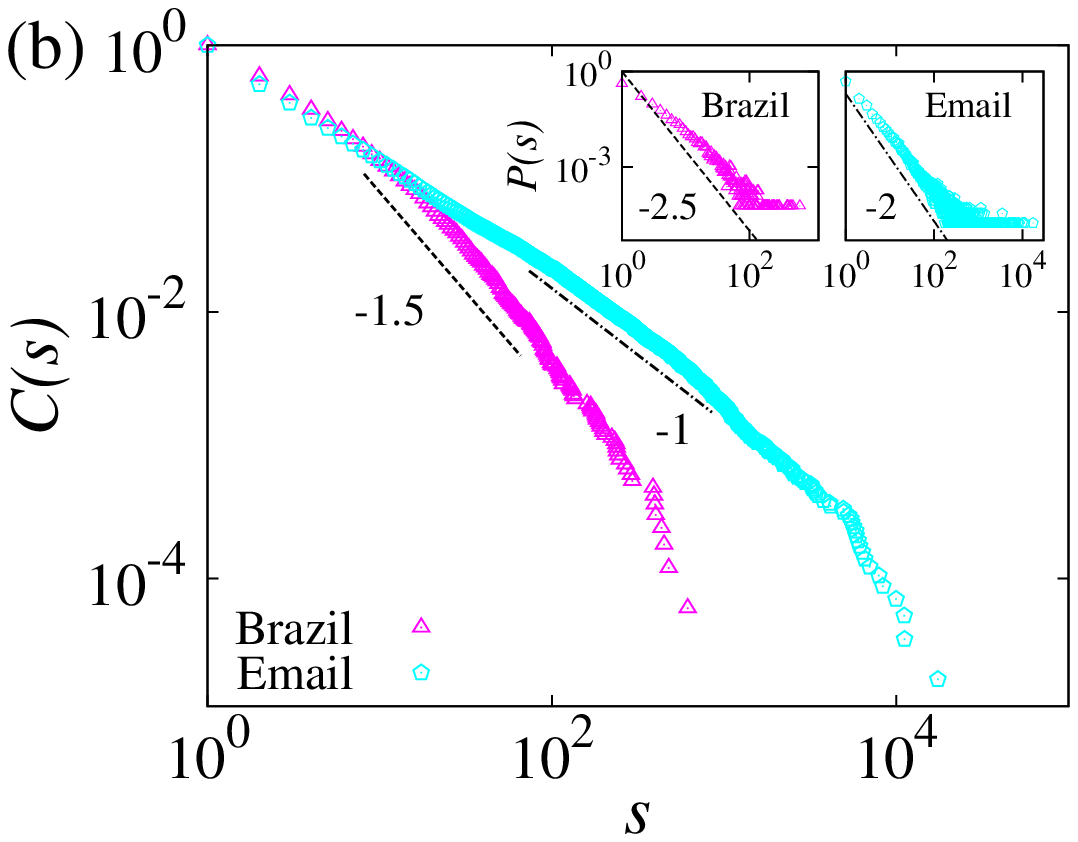}
\caption{At $t_w=T$, the complementary cumulative strength distribution, $C(s)$, is measured for five datasets, respectively: (a) School (2011, 2012) and Hospital cases decay exponentially with slightly different (almost homogeneous) strengths, while (b) Brazil and Email cases decay as power-law with heterogeneous strengths. In the insets of (b), the strength distribution, $P(s)$, is shown as well.}
\label{fig2CCDF}       
\end{figure}

\subsection{Activity and memory in empirical data}
\label{model-setup}

We define the activity rate at a node $i$ as $a_i$ (the number of appearance per time), which is directly related to the strength, $s_i=ma_iT+{\ell}_i(T)$, in static networks. Here ${\ell}_i(T)$ is the number of connections from other active nodes until the maximum time $t=T$, and $m$ is the number of links generated from an active node. It is noted that the strength distribution in temporal networks, can be obtained from the time-aggregated representation at $t=t_w$. 

Figure~\ref{fig2CCDF} shows how the strength distribution behaves, $P(s)\sim s^{-\gamma}\exp(-s/c)$, and its complementary cumulative function, $C(s)$, decays for five empirical datasets. Note that the functional shape indicates the heterogeneity of nodes' activities. Such results can be categorized into the two classes of the activity distribution $P(a)$ as follows:
\begin{equation}
P(a)\sim a^{-\gamma}\exp(-a/c)
\\
\sim\left\{
\begin{array}{lr}
\exp(-a/c) \\
a^{-\gamma}
\end{array}
\begin{array}{ll}
& (\textrm{exponential}), \\ 
& (\textrm{power-law}),
\end{array}
\right.
\end{equation}
where $\gamma$ is the activity exponent with $\gamma\ge0$, and $c$ is the characteristic size of activity when $\gamma=0$.  

\begin{figure*}[]
\center
\includegraphics[width=0.32\textwidth]{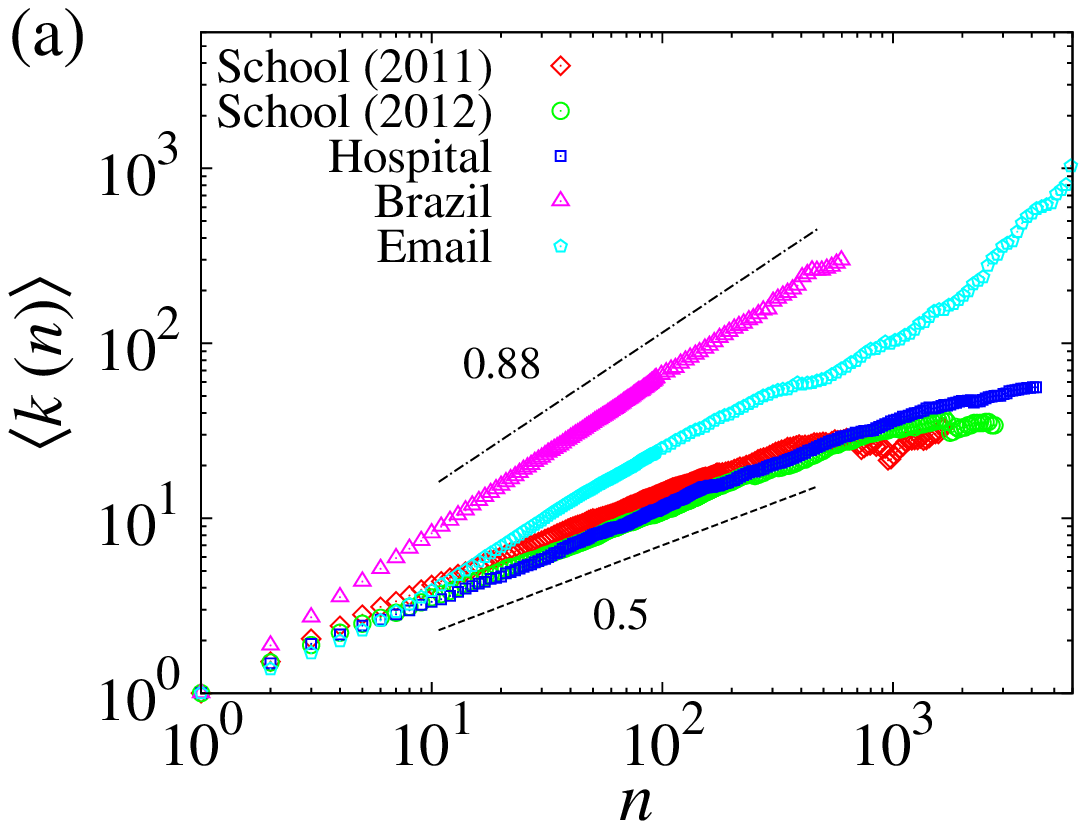}
\includegraphics[width=0.32\textwidth]{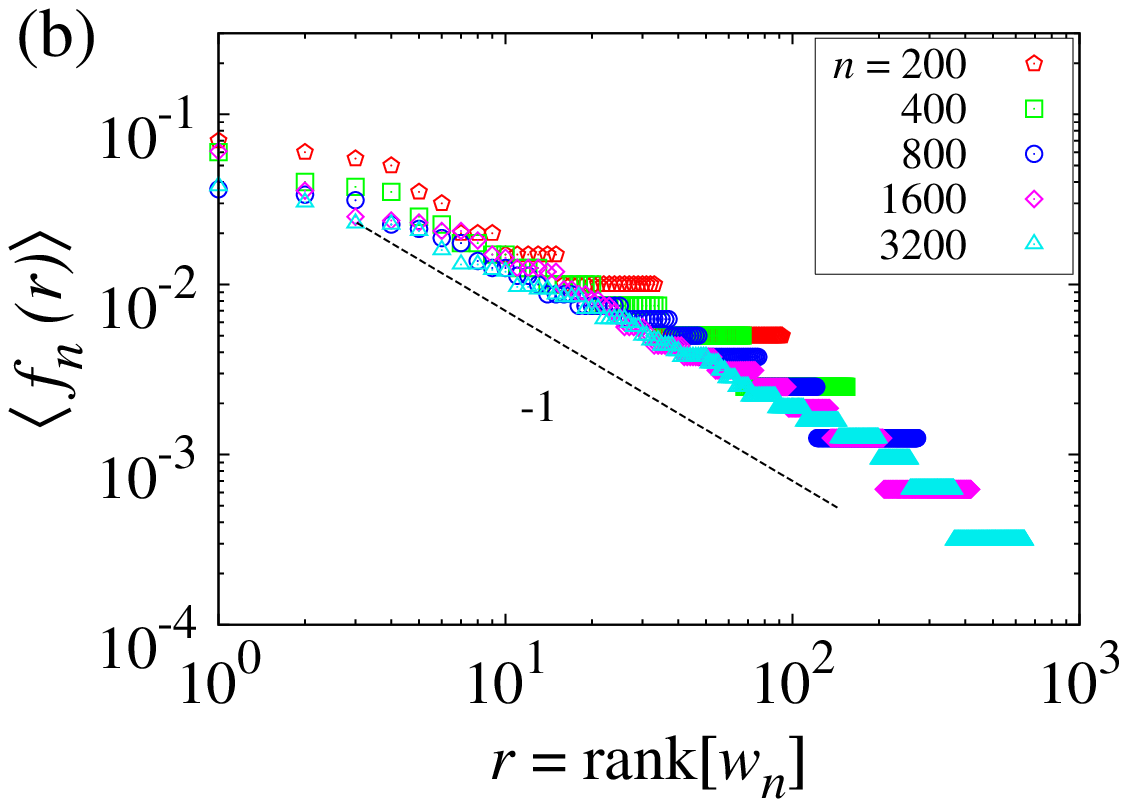}
\includegraphics[width=0.33\textwidth]{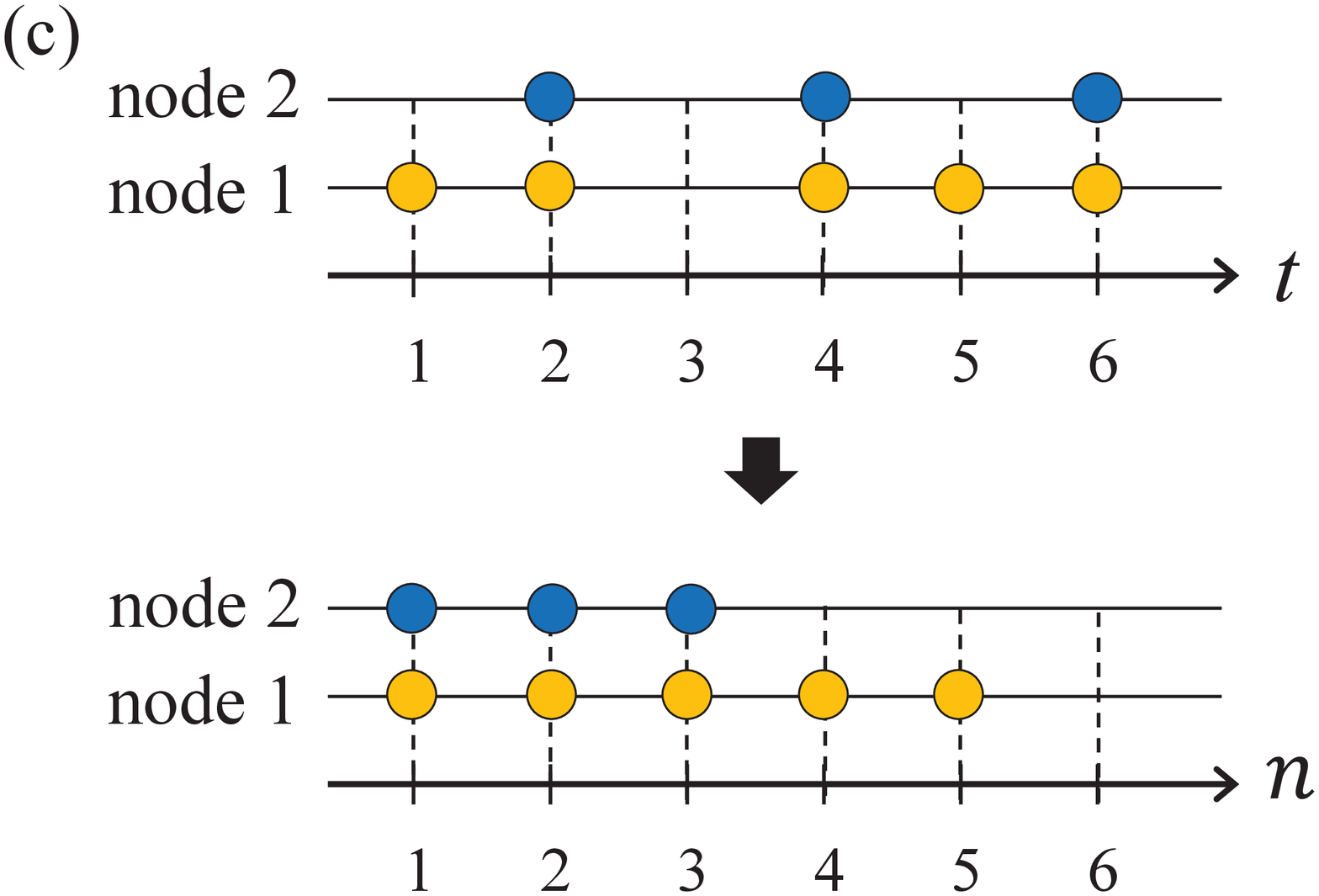}
\caption{For five datasets (see Tab.~\ref{tab-data}) that are considered as time-aggregated networks: (a) the average degree, $\langle k(n)\rangle$, is plotted against the personal contact-time step, $n$, which scales as $\langle k(n)\rangle\sim n^{\alpha}$ with $0<\alpha<1.$; (b) the normalized rank frequency, $\langle f_n(r)\rangle$, is plotted against the rank of the link-weight list ($\{w_n\}$), $r={\rm rank}[w_n]$, for various $n=200,~400,~...,3200$; (c) For a better understanding of $n$, the comparison of $n$ with real time $t$ are illustrated. In the static representation, $n_{i,{\rm max}}=s_i.$ }
\label{fig3pattern}       
\end{figure*}
\begin{figure*}[]
\center
\includegraphics*[width=\textwidth]{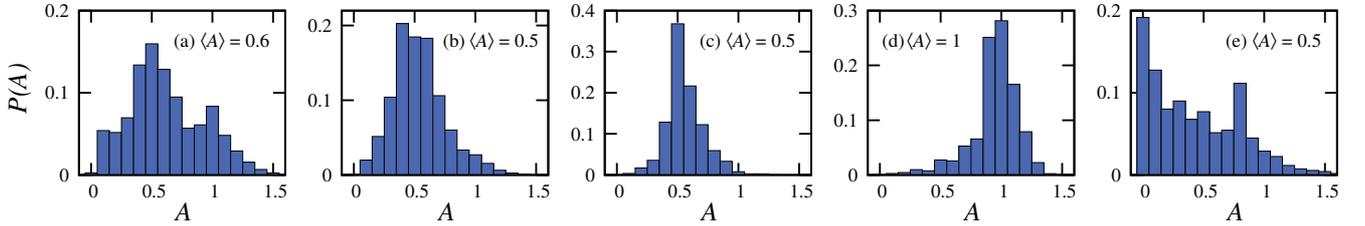}
\caption{The histograms of the memory coefficient $A$ are measured in five empirical datasets: (a) School (2011) (b) School (2012) (c) Hospital (d) Brazil (e) Email, respectively, where the average value of $A$ is shown inside each plot.}
\label{fig4A}       
\end{figure*}
\begin{table}[]
\caption{All the detailed information for five datasets are summarized, where $N$ is the total number of nodes, $T$ is the maximum time (the total number of interaction steps), $\gamma$ is the strength/activity exponent, $c$ is the characteristic strength/activity, $A$ is the average memory coefficient, and $\beta$ is the memory exponent.}
\label{tab-data}
\begin{tabular}{ccccccc}
\hline\hline
Dataset & $N$ & $T$ (step) & $\gamma$ & $c$ & $A$ & $\beta$ \\
\hline\hline
School (2011) & 126 & 28~561 & 0 & $400$ & 0.6 & 0.50 \\
School (2012) & 180 & 45~047 & 0 & $500$ & 0.5 & 0.50 \\
Hospital & 75 & 32~424 & 0 & $1000$ & 0.5 & 0.50 \\
Brazil & 16~730 & 50~632 & 2.5 & $\infty$ & 1.0 & 0.12 \\
Email & 57~182 & 447~543 & 2.0 & $\infty$ & 0.5 & 0.20 \\
\hline\hline
\end{tabular}
\end{table}
\begin{figure*}[]
\center
\begin{tabular}{c|c}
\includegraphics*[width=0.425\textwidth]{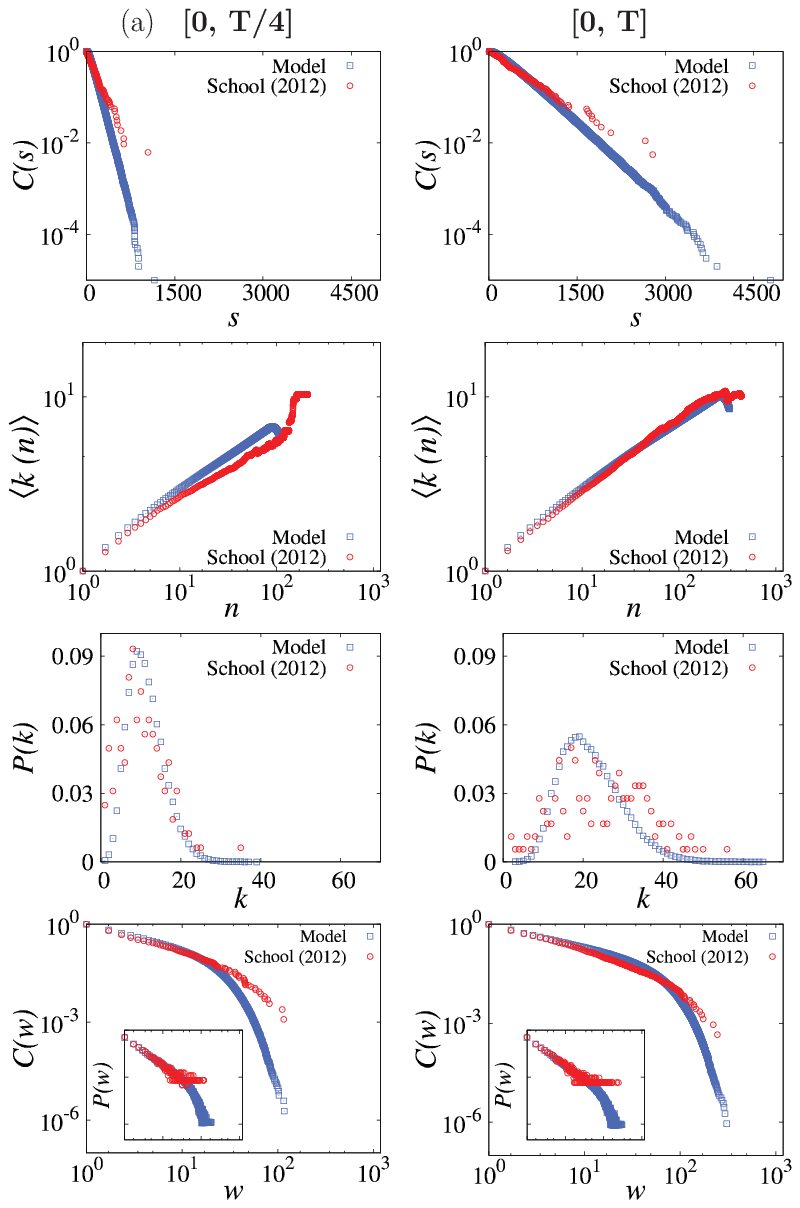}&
\includegraphics*[width=0.425\textwidth]{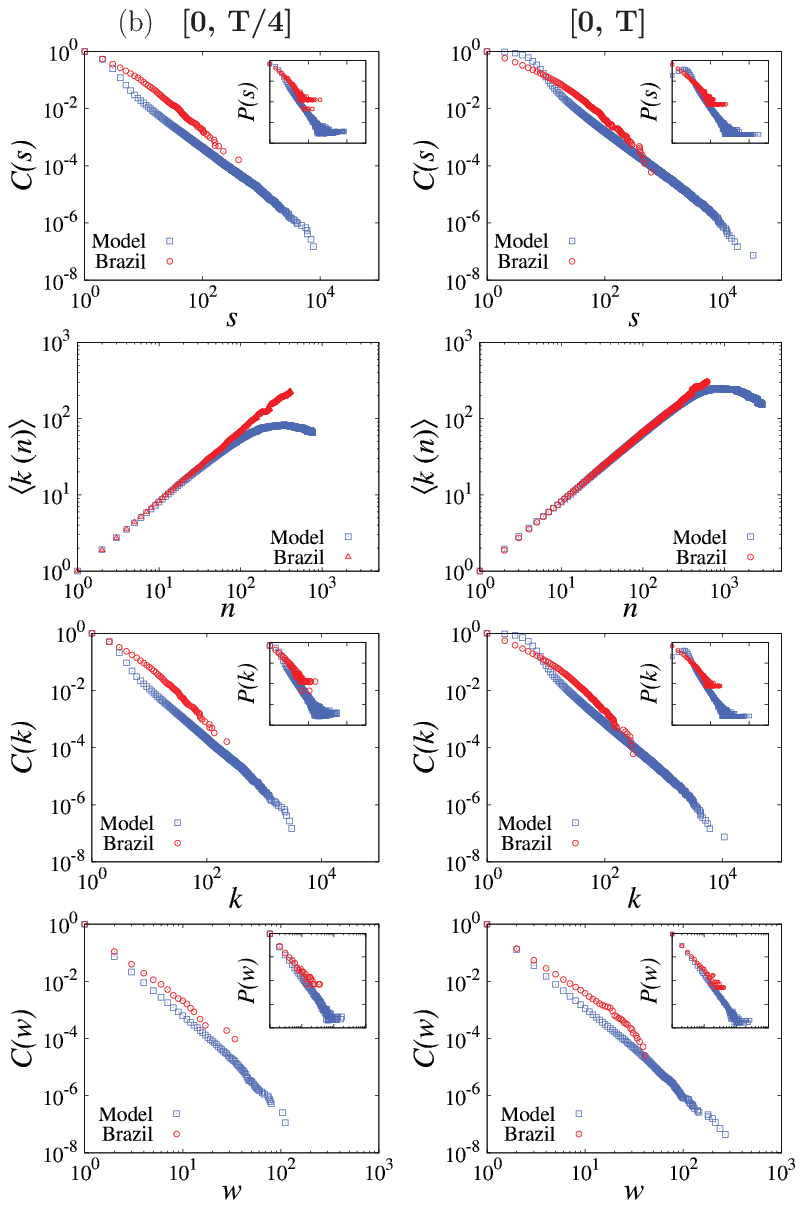}
\end{tabular}
\caption{Network properties for two datasets, are compared to those for the model networks at $t_w=T/4$ and $t_w=T$, respectively: (a) School (2012) vs. the model with $\{N=100,~T=25~000~{\rm (step)},~\gamma=0,~c=1/500,~A=0.5,~\beta=0.50\}$. (b) Brazil vs. the model with $\{N=15~000,~T=50~000~{\rm (step)},~\gamma=2.5,~c=\infty,~A=1.0,~\beta=0.12\}$.}
\label{fig5model-test}
\end{figure*}

In contrast, the memory effect can be described as the contact pattern and preference between nodes in time-varying networks. As shown in Figure~\ref{fig3pattern}, we observe that the tendency to create new links are not random when the memory exists, where we measure two following quantities with personal contact-time step, $n$ (not real time $t$), see the difference between $t$ and $n$ in Figure~\ref{fig3pattern}c: (i) the average degree as a function of $n$, $\langle k(n)\rangle$, and (ii) the normalized rank frequency, $\langle f_n(r)\rangle$, against the rank of weighted links ($\{w_n\}$), $r={\rm rank}[w_n]$. When each node can memorize its past, we find that the tendency to contact previously connected nodes with the different link preference, as $n$ increases, which is similar to the scaling patterns of visited locations in human mobility~\cite{Song2010}. 

In Figure~\ref{fig3pattern}a, we observe that $\langle k(n)\rangle$ scales as  
\begin{equation}
\langle k(n)\rangle=\frac{\sum_{i=1}^{N}k_i\delta(n_i-n)}{\sum_{i=1}^{N}\delta(n_i-n)}\sim n^{\alpha}.
\end{equation}
where $k_i$ is the cumulative degree of a node $i$ up to the contact-time step $n_i$, $\delta(x)$ is the Dirac delta function, and $N$ is the total number of nodes in a time-aggregated network at $t_w=T$. 

As a result, the probability to contact a new one is described as follows:
\begin{equation} 
P_{\rm new}(n)=\frac{{\rm d}\langle k(n)\rangle}{{\rm d}n}=An^{\alpha-1}=An^{-\beta},
\end{equation}
where $\beta$ is the memory exponent with $0\le \beta\le 1$, and $A$ is the memory coefficient. In other words, nodes interact with previously connected nodes with the probability, $P_{\rm old}(n)=1-P_{\rm new}(n)$, and the memory effect gets stronger as $\beta\to 1$. Here we set the memory coefficient $A$ to the average value of memory coefficient $\langle A\rangle$, which is taken from the histograms of $A$ in five empirical datasets (see Fig.~\ref{fig4A}).

In Figure~\ref{fig3pattern}b, we find that the different preference exists in the choice of ever connected neighbors per node, where we measure the normalized rank frequency of weight links, $\langle f_n(r)\rangle$: 
\begin{equation}
\langle f_n(r)\rangle
=\frac{\sum_{i=1}^{N}f_i(r)\theta(n_i-n+1)}{\sum_{i=1}^{N}\theta(n_i-n+1)}\sim r^{-1},
\end{equation}
where $f_i(r)\equiv {\rm rank}[\{w_{il}\}]/n_i$ at a node $i$, and $\theta(x)$ is a Heaviside step function.
It implies that nodes have a propensity to interact with the nodes that are frequently connected before.

\subsection{Model network formation}
\label{model-network}

Based on the results obtained from five empirical datasets in the time-aggregated representation, we propose a simple network model of $N$ nodes, which is designed to capture the emergence of heterogeneous network properties with the memory effect as follows:
\begin{enumerate}
\item{First, consider $N$ disconnected nodes.}
\item{At each discrete time increment $\Delta t$, each node can be active with the probability $p_i=a_i/\max_{i=1}^{N}\{a_i\}$. Once a node $i$ becomes active, it can generate $m$ links.}
\item{Each active node $i$ follows: 
\begin{itemize}
\item{Case 1 - If the activation is for the first time, connect the node $i$ to any node at random, except for itself.} 
\item{Case 2 - Otherwise, do either (i) or (ii):   
\begin{itemize}
\item{(i) Create a new link with the probability $$P_{{\rm new},i}=An_i^{-\beta},$$ and connect it to a new node at random.} 
\item{(ii) With $P_{{\rm old},i}=1-P_{{\rm new},i}$, connect the node $i$ to one of ever connected nodes with the probability $$\Pi_{ij}(n_i)=\frac{w_{ij}(n_i)}{\sum_{l=1}^{k_i}w_{il}(n_i)}$$.}
\end{itemize}}
\end{itemize}}
\item{At the next time step $t+{\Delta}t$, update the memory for every node, and delete all the edges in the network at $t$.}
\item{Repeat the above procedure until $t=T$.}
\end{enumerate}
Here all the parameters in the model are based on five datasets (see Tab.~\ref{tab-data}), and we set $m=1$ for short-time scale interactions without loss of generality.

\subsection{Network analysis: Model vs. real data}
\label{model-detail}

For a better understanding of two selected datasets, School (2012) and Brazil, we perform numerical simulations in terms of two model settings based on Table~\ref{tab-data}. 

In Figures~\ref{fig5model-test}a and 5b, network properties for the model are compared with those for the corresponding dataset at $t_w=T/4$ and $t_w=T$, respectively: (a) School (2012) and (b) Brazil. Although all the other correlations are ignored, they exhibit pretty similar behaviors. Models reproduce quite well not only in the strength distribution and the average degree of the time-dependent strength, but also in the degree distribution and the link-weight distribution. It implies that the model mechanism  enables us to explain the procedure of interactions in temporal networks.

\begin{SCfigure*}[]
\includegraphics[width=0.32\textwidth]{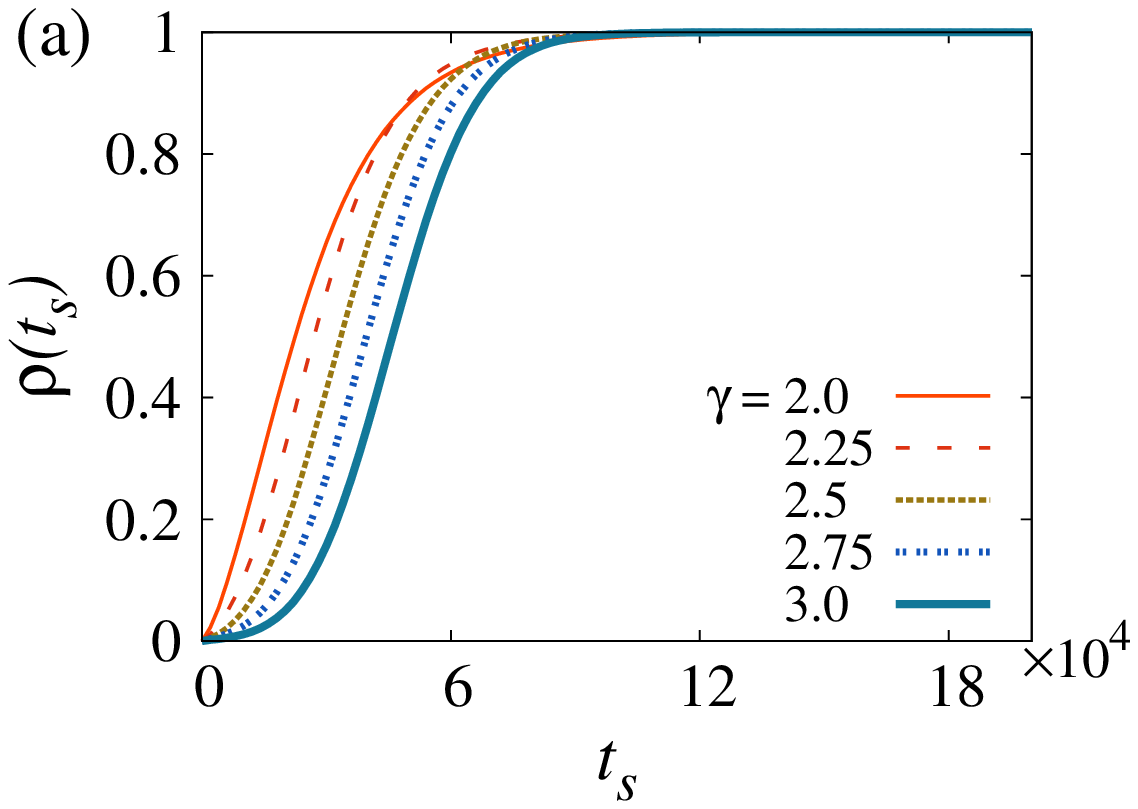}
\includegraphics[width=0.32\textwidth]{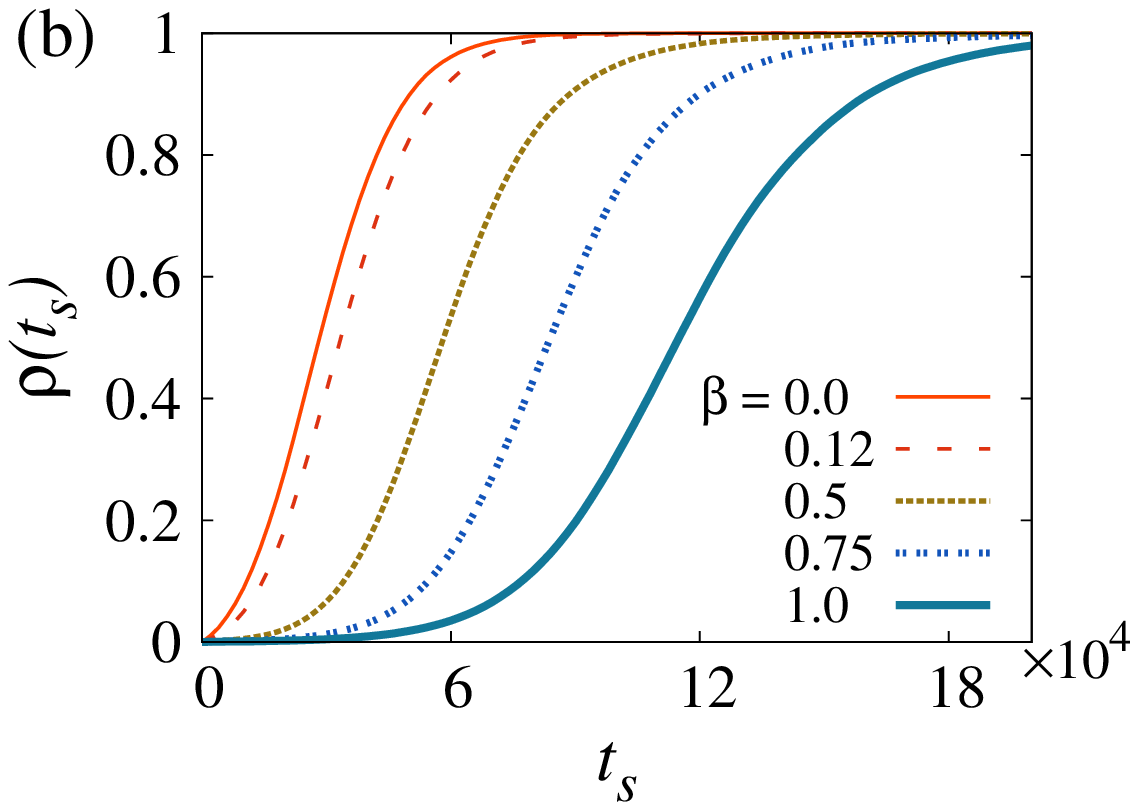}
\caption{For Brazilian dataset-like time-aggregated networks of $N=15~000$ and $T=200~000$ (step), we investigate the relevance of activity and memory in epidemic spreading, where we measure the infected density ($\rho$) as time step  ($t_s$) increases in the SI: (a) $\beta=0.12$ for various activity-exponent values, $\gamma=2.0,...,3.0$ and (b) $\gamma=2.5$ for various memory-exponent values, $\beta=0.0,...,1.0$.}
\label{fig6SI-1}       
\end{SCfigure*}
\begin{figure*}[]
\center
\includegraphics[width=0.32\textwidth]{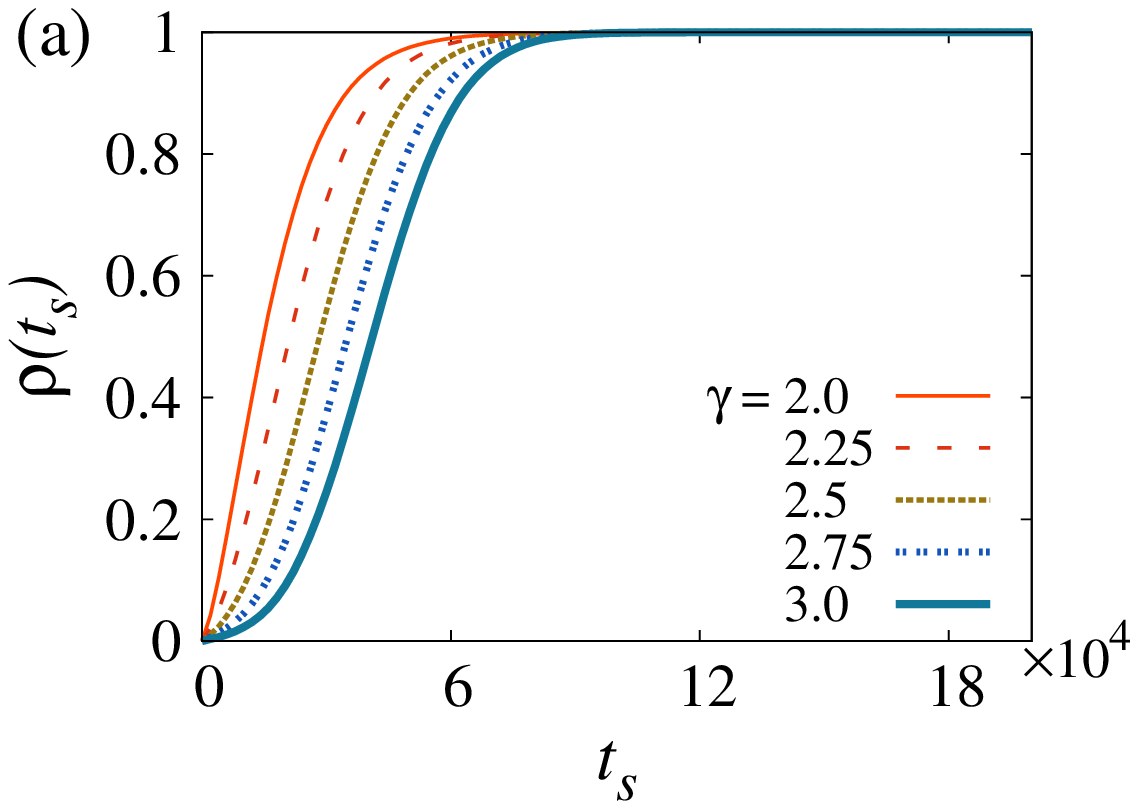}
\includegraphics[width=0.32\textwidth]{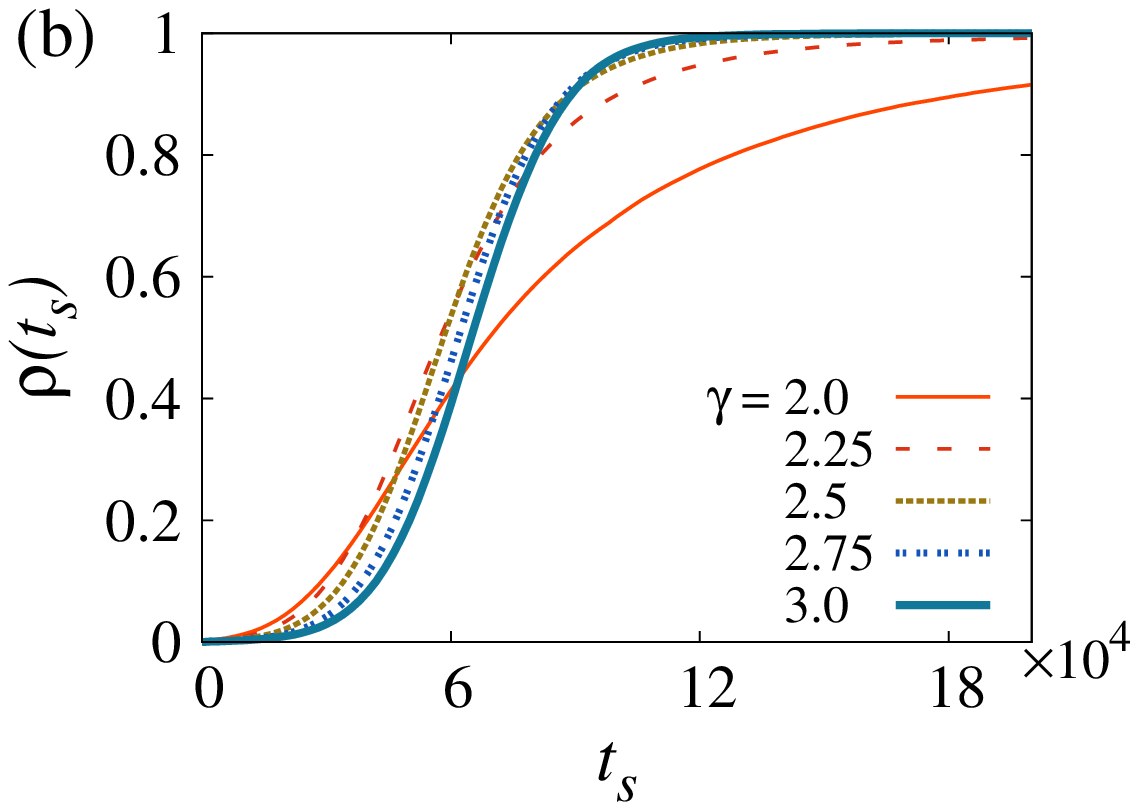}
\includegraphics[width=0.32\textwidth]{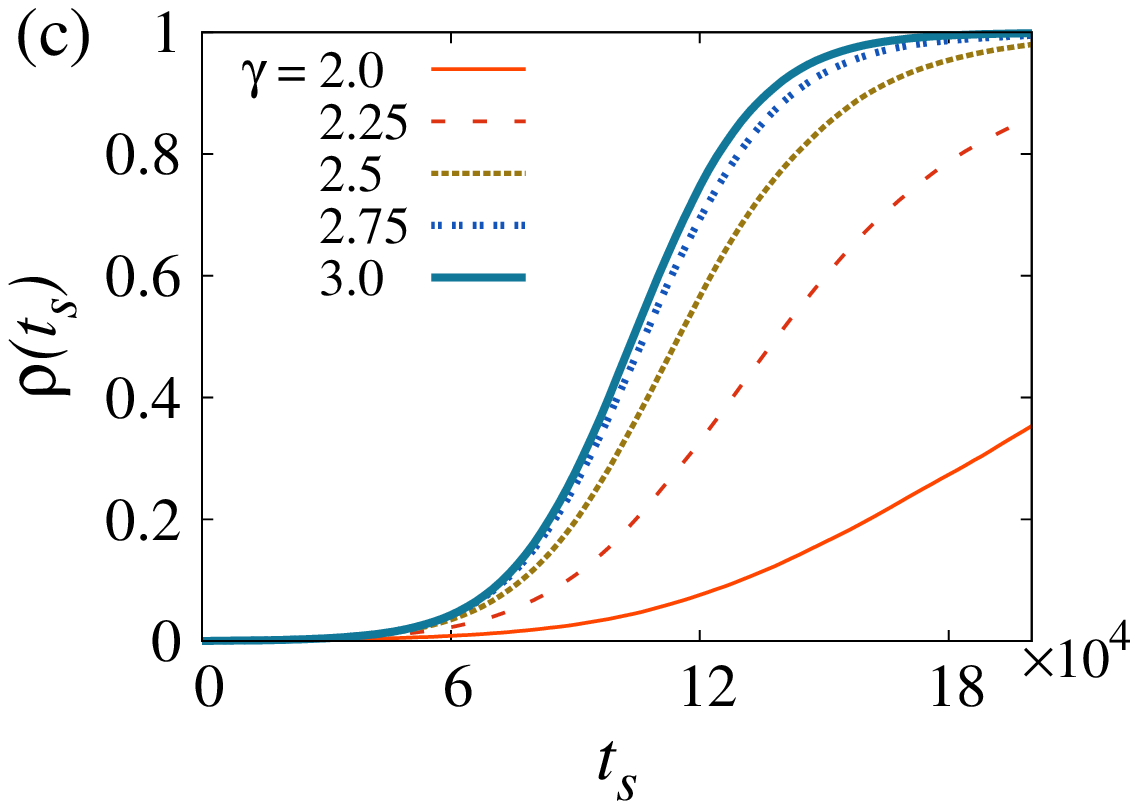}
\caption{The role of activity in the SI at the three values of the memory exponent: $\beta=0~(a),~0.5~(b)$ and $1.0~(c)$, where other network settings are exactly the same as those in Figure~\ref{fig6SI-1}. It is found that the memory effect seems to be stronger in the networks  with highly active nodes, so that the order of saturation patterns is changed.}
\label{fig7SI-2}       
\end{figure*}

\section{Numerical Results}
\label{numerics}

\begin{figure}[]
\center
\includegraphics[width=0.75\columnwidth]{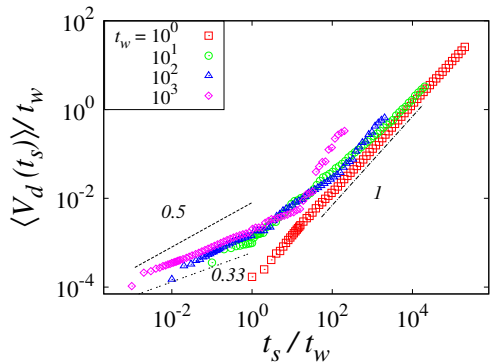}
\caption{In time-aggregated networks at the size of time window, $t_w$, the average number of distinct nodes visited by a random walker, $\langle V_d(t_s)\rangle$, by a single greedy random walker, can reveal the network formation procedure as well as the steady-state topology as $t_s$ elapses, starting at the aging time $t_a=0$.}
\label{fig8RW}       
\end{figure}

In this section, we investigate the role of activity and memory in epidemics for time-aggregated networks, the temporal-pattern coarsening effect on the diffusion by a random greedy walker in time-varying networks, and the aging effect on temporal coarsening schemes, where we employ the extended finite-size scaling (FSS) technique in terms of two time scales: One is the aging time (when the observation starts) $t_a$, and the other is the time-window size $t_w$, and $N$. In particular, we measure the time evolution of physical quantities in terms of the time step: $t_s$ (the sequence index of the link addition for all nodes) for Figures~\ref{fig6SI-1}-\ref{fig8RW}; the real time $t$ [arbitrary unit] (a.u.) for Figures~\ref{fig9aging-1} and~\ref{fig10aging-2}.

\subsection{Role of activity and memory in epidemic spreading}

We revisit the conventional susceptible-infected model (SI) to investigate how activity and memory affect epidemic spreading processes. Due to the fact that the SI is the simplest model of epidemics, lots of network studies have discussed about the epidemic spreading behavior using the SI and its variants~(see the most recent review of epidemic spreading on complex networks~\cite{EP-RMP2015}). Hence it is good to be used for the first test of epidemic spreading in our model. 

In particular, we focus on how the activity plays a role in the time evolution of the infection density, $\rho(t_s)=I(t_s)/N$, compared to how the memory effect does, where we set at the infection probability $\lambda=1$. In the SI with $\lambda=1$, a health node gets infected once it has at least one infected node among its neighors on model networks that are controlled by the activity exponent $\gamma$ from $P(a)\sim a^{-\gamma}$ and the memory exponent $\beta$ from $P_{\rm new}(n)=An^{-\beta}$. For homogeneous activities, it is found that even the strongest memory effect ($\beta=1$) does not alter the behaviors much, so we here only consider both in power-law cases.   

In Figures~\ref{fig6SI-1} and~\ref{fig7SI-2}, we observe how $\rho(t_s)=I(t_s)/N$ approaches one as $t_s$ (time step) elapses. In particular, we investigate the role of activity and memory in epidemic spreading, respectively, for various $\gamma$ at a fixed $\beta=0.12$ and for various $\beta$ at a fixed $\gamma=2.5$. The speed of spreading increases as $\gamma\to 2$ or $\beta\to 0$ (see Fig. \ref{fig6SI-1}). These results indicate that the networks composed of highly active or weak memory nodes have the higher probability of contacting healthy nodes, so that they have the faster speed of spreading. Such optimized spreading phenomena are observed in Figure~\ref{fig7SI-2}, where various memory settings are tested with the same activity distribution. The extended FSS analysis of the SI is still ongoing as a separated project with the generalized version. 

\subsection{Temporal coarsening tests on dynamics}

The difference between temporal and static networks is speculated to understand the origin of the optimized spreading against the power-law types of activities and memories, where we consider a random greedy walker, and let the random walker diffuse on model networks as $t_s$ for the size of time window, $t_w$, varies. Through the history of visiting nodes for the single walker, we measure the average number of distinct nodes visited by the walker, $\langle V_d(t_s)\rangle$. It provides the following extended FSS form as $t_s$ varies for various $t_w$ because temporal coarsening effectively changes the network size at $t_w$ as $N_w(N,t_w)$: 
\begin{equation}
\langle V_d(t_s)\rangle=t_w f(t_s/t_w)
\sim\left\{
\begin{array}{lr}
t_s\\
t_s^{\zeta}t_w^{1-\zeta}
\end{array}
\begin{array}{ll}
& (\textrm{for}~t_s>t_w), \\ 
& (\textrm{for}~t_s<t_w),
\end{array}
\right.
\end{equation}
where $f(x)\sim x$ for $x>1$; $x^{\zeta}$ with $\zeta<1$ for $x<1$.

As shown in Figure~\ref{fig8RW}, data of $\langle V_d(t_s)\rangle/t_w$ collapse quite well, as a function of $x=t_s/t_w$, where we use the Brazilian model network with the setting of $\gamma=2.5$ and $\beta=0.12$ (exactly the same as Fig.~\ref{fig5model-test}). Linear scaling behaviors here would be rather trivial if there is either memoryless or multiple links, while for $t_s<t_w$, $\langle V_d(t_s)\rangle/t_w\sim (t_s/t_w)^{\zeta}$ with with $1/3\le \zeta\le 1/2$. For the latter, the detailed discussion is necessary as a future work. In the stage of writing our manuscript and in process of submission, we find that the similar work was posted by Saram{\"a}ki and Holme~\cite{SHolme2015}, where greedy walks in temporal networks are discussed as a tool for exploring and probing temporal-topological structures without the detailed discussion of scaling behaviors. As a future study, it would be quite interesting if scaling properties are carefully discussed in the same context.

\begin{figure}[]
\center
\includegraphics[width=0.75\columnwidth]{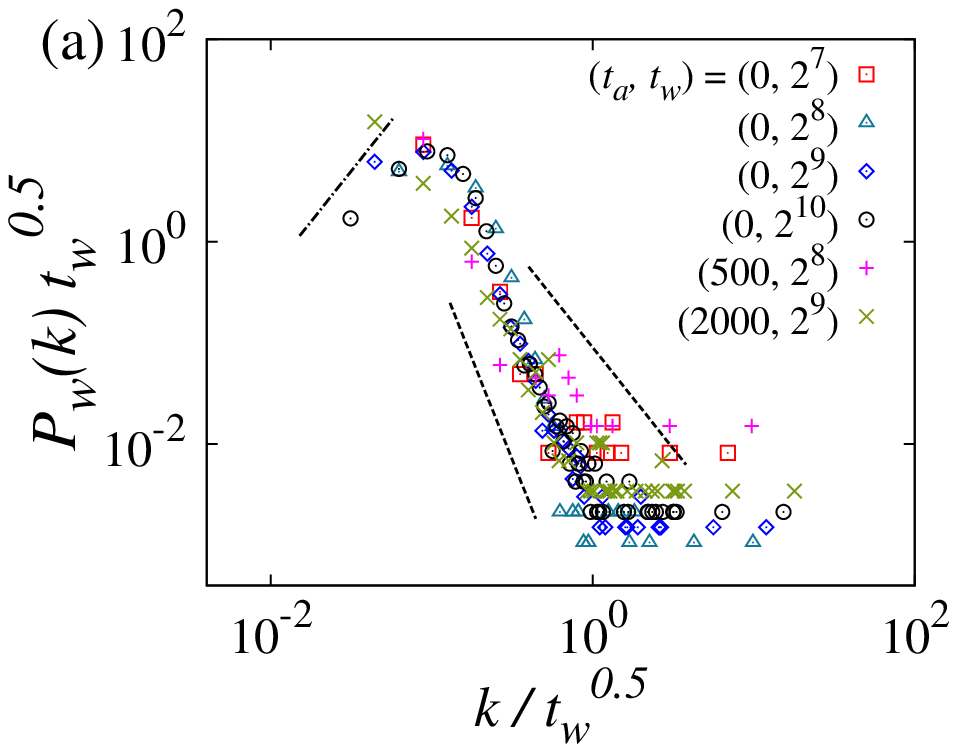}\\
\includegraphics[width=0.75\columnwidth]{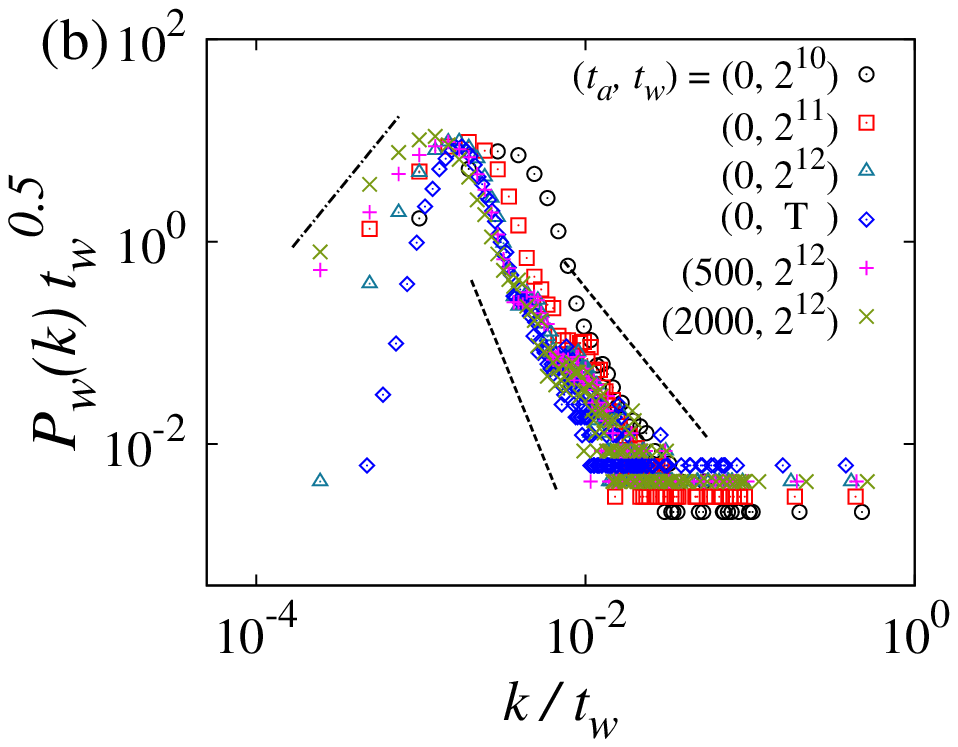}
\caption{Scaling collapse of $P_{w}(k)t_w^{0.5}$, versus the rescaled degree, $k/t_w^{\eta}$ for various $t_w$: (a) $\eta=0.5$ in the regime that the average number of nodes, $\langle N_w(N,t_w)\rangle_{a}$, increases, at the aging time $t_a=0,~500,~2000~{\rm (a.u.)}$ for various $t_w<t_{\rm sat}$, while (b) $\eta=1$ in the regime that $\langle N_w(N,t_w)\rangle_{a}$ gets saturated at $t_a=0,~500,~2000~{\rm (a.u.)}$ for large $t_w>t_{\rm sat}$. Here $t_{\rm sat}\simeq 10^{3}\sim 2^{10}$ for $N=15~000$ and the time scale is taken with real time (a.u.), $T=8496~{\rm (a.u.)~or}~200~000~{\rm (step)}$. The slopes of guided lines are 2.0, -4.0 and -2.0 (from left to right), respectively.}
\label{fig9aging-1}       
\end{figure}

\begin{figure*}[]
\center
\includegraphics[width=0.475\textwidth]{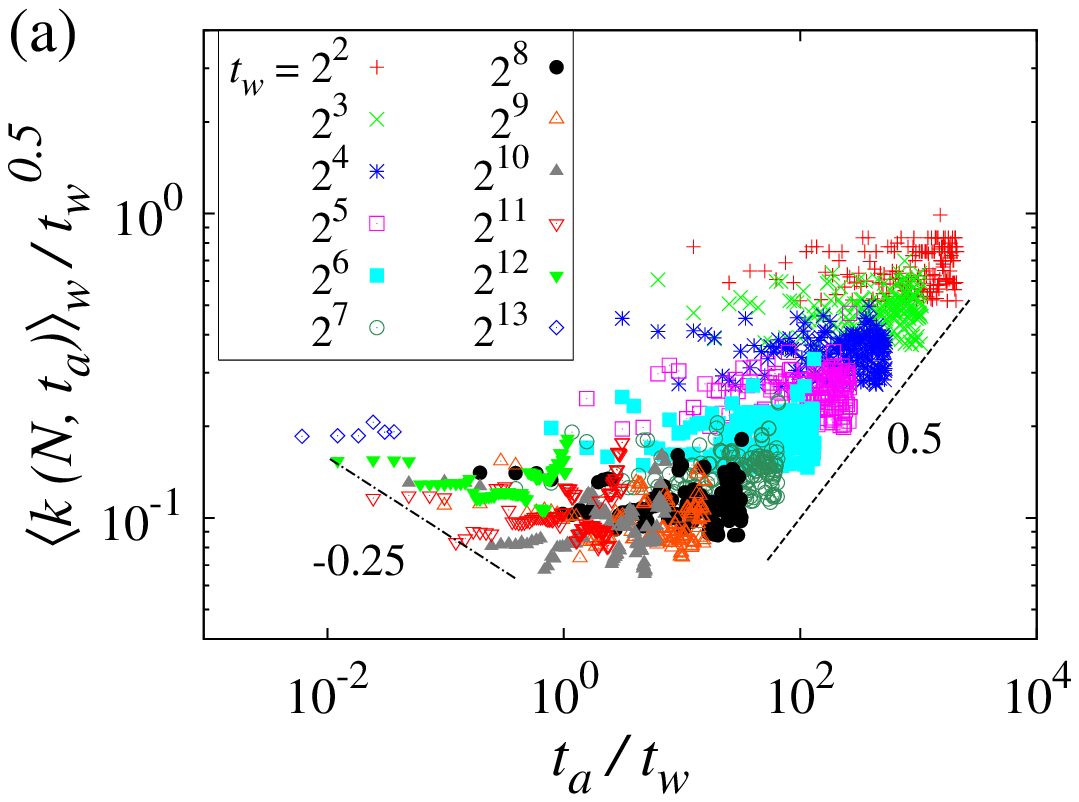}
\includegraphics[width=0.475\textwidth]{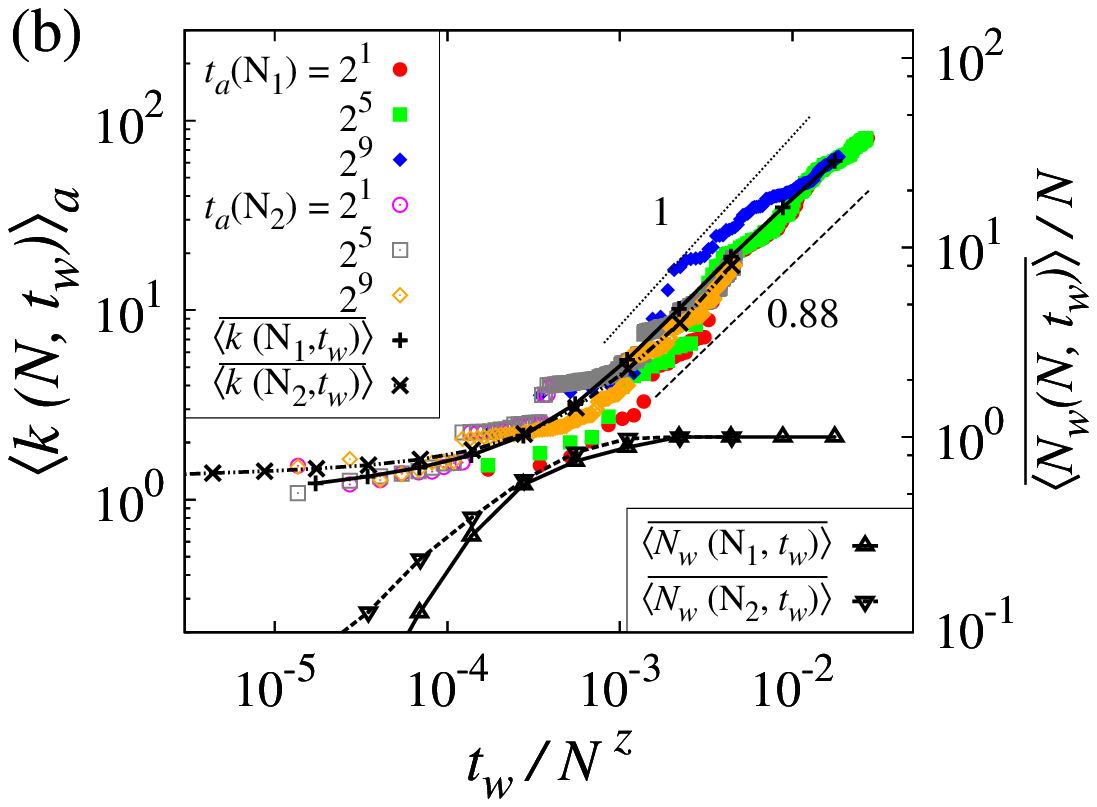}
\caption{Scaling behaviors of the average degree are plotted as a function of $t_w,~t_a,~{\rm and}~N$: (a) $\langle \langle k(N,t_a)\rangle_{w}/t_w^{0.5}$ versus $t_a/t_w$ for various $t_w$, and (b) $\langle k(N,t_w)\rangle_{a}$ versus $t_w/N^z$ with $z=1.5$, where $\overline{\langle k(N,t_w)\rangle}_{a}$ and $\overline{\langle N_w(N,t_w)\rangle}/N$ scale as the same variable $x=t_w/N^{z}$. Here $\langle {\cal O}\rangle$ is the network ensemble average and $\overline{\cal O}$ is the temporal average for $N_1=1500$ and $N_2=10N_1$.}
\label{fig10aging-2}       
\end{figure*}

\subsection{Aging tests in temporal coarsening}

In addition, we study the detailed structures of model networks as the aging time $t_a$ and the observation time-window size $t_w$ vary using aging/history effects. Most recently, burstiness and aging have been discussed in temporal networks~\cite{Moinet2015}, which presented how to check the presence of aging effects and bursting behaviors. Aging effects are performed on model networks for the dataset of Brazil. 
As expected, neither aging effects nor bursting are observed. However, we investigate the extended FSS properties for $t_w$ and $N$ (see Figs.~\ref{fig9aging-1} and~\ref{fig10aging-2}). 

Figure~\ref{fig9aging-1} exhibits how the degree distribution $P_w(k)$ decays as $k$ at $t_w$ for various $t_a$. While $t_a$ is irrelevant to scaling behaviors, $t_w$ is found to be relevant to those. It is because the total number of nodes, $N$, is constant, and the effective number of active nodes at $t_w$, $N_w(N,t_w)$, is almost linear to $t_w$. As $t_w$ gets wider enough to cover the total number of nodes, $N$, irrespective of $t_a$, $P_{w}(k)$ becomes changed from power-law decay to bell-shaped: 
\begin{equation} 
P_{w}(k)=t_w^{-0.5}F(k/t_w^{\eta}),
\end{equation}     
where $F(x)\sim x^{-\gamma_k}$ with $2<\gamma_k<4$ for $x>1$; $x^2$ for $x<1$, and
where the most probable degree $k_{\rm mp}\sim t_w^{\eta}$ with $\eta=0.5\to 1$. Note that the shape of the degree distribution has been briefly reported in the earlier work~\cite{Memory-JSTAT2014} with and without node population growth.

Figure~\ref{fig10aging-2} provides the basic analysis of time-aggregated network topologies in terms of the scaling behaviors of the average degree for various $t_a$ and $t_w$. In Figure~\ref{fig10aging-2}a, it is observed again that $t_a$ is irrelevant to the scaling behavior of $\langle k(N, t_a)\rangle_{w}$ but it depends on $t_w$ as follows:
\begin{equation}
\langle k(N,t_a)\rangle_{w}=\sum_{i=1}^{N}k_i(t_a)/N_a(N,t_a)=t_w^{0.5}g(t_a/t_w),
\end{equation} 
where $N_a(N,t_a)=\sum_{i=1}^{N}\theta[k_i(t_a)-1]$ and $\theta(x)=1$ only if $x>0$ (0 if $x<0$). Moreover, we find that $g(x)=x^{-0.25}$ for $x\ll 1$; constant for $x\sim 1$; $x^{0.5}$ for $x\gg 1$. In Figure~\ref{fig10aging-2}b, we also investigate the temporal average of $\langle N_w(N,t_w)\rangle_{a}$ and $\langle k(N,t_w)\rangle_{a}$, which are defined as
\begin{equation}
\overline{\langle {\cal O}(N,t_w)\rangle}=
\sum_{\tau=t_a}^{T-t_w}\langle {\cal O}(N,t_w)\rangle_{\tau}/(T-t_w-t_a).
\end{equation}
The results from Figure~\ref{fig10aging-2} tell us that the average degree linearly scales as $x=t_w/N^{z}$ with $z=1.5(>1)$ for $x\gg 10^{-3}$, while the average fraction of active nodes linearly scales as $x$ for $x\ll 10^{-3}$.

\section{Summary and remarks}
\label{summary}

We have analyzed both the interaction pattern in the time-aggregated network representation, and the contact pattern of empirical temporal networks. In particular, we proposed the simplest model of network formation procedures for time-varying networks with memory, which is the best combination of the existing models with activity and memory in some better aspects. The generation of model networks enables us to express the nature of interaction for short-time scales, unlike connectivity-driven models which capture mostly static network structures. 

In our model, both activity and memory are denoted by the individual characteristics without the other correlations, e.g., the bursting behavior and the variation of link densities with time. Nevertheless, the model network is well-suited for capturing the scaling behaviors of cumulative distributions, the process of the interaction. Such a simple mechanism may be useful to take a quantitative approach, e.g., scaling relations among the exponents and critical behaviors at the epidemic threshold of epidemic models. Moreover, we observed how activity and memory play a role in time-varying networks for various model setting parameters, where we quantified such effects, tested on spreading dynamics by the SI. For a better understanding of the difference between temporal and static networks, we consider the temporal coarsening by using various time-window ($t_w$) sizes, we tested the diffusion by a random greedy walker and the aging effect caused by the existence of burstiness. 

Finally, the temporal-coarsening pattern analysis as well as the optimal size of observation-time window ($t_w$) should be carefully discussed as a separated topic (similar to Ref.~\cite{Caceres2011}) in elsewhere since there are lots of important issues and open problems to be speculated in the context of the universality class and the validity for the extended FSS properties.
    
\section*{Author contribution statement}
All authors designed the research and wrote the manuscript equally. In particular, H.K. and M.H. developed analytic results with intuitive arguments, and conducted numerical simulations.\\

\noindent{This work was supported by the National Research Foundation of Korea (NRF) by the Korea government [Grants No. 2014R1A1A4A01003864 (H.K., M.H.) and No. 2011-0028908 (H.K., H.J.)].}

%
\bibliography{EPJB2015-resub}
%
%
%

\end{document}